\input epsf
\documentstyle[preprint,aps,floats,amsfonts]{revtex}
\begin{document}
\draft
\title{Non-supersymmetric cousins of supersymmetric gauge
theories:\protect\\  quantum space of parameters and double scaling limits}
\author{Frank Ferrari}
\address{Joseph Henry Laboratories, Princeton University, Princeton, New
Jersey 08544, USA \\
and Centre National de la Recherche Scientifique, \'Ecole Normale
Sup\'erieure, Paris, France\\
{\tt fferrari@feynman.princeton.edu}}
\date{March 2000}
\maketitle
\setbox1=\hbox{PUPT-1918}
\setbox2=\hbox{LPTENS-00/05}
\setbox3=\hbox{\tt hep-th@xxx/0003142}
\makeatletter
\global\@specialpagefalse
\def\@oddhead{\hfill\vbox{\box1\box2\box3}}
\let\@evenhead\@oddhead
\begin{abstract}
I point out that standard two dimensional, asymptotically free,
non-linear sigma models, supplemented with terms giving a mass to 
the would-be Goldstone bosons, share many properties with four
dimensional supersymmetric gauge theories, and are tractable even in the
non-supersymmetric cases. The space of mass parameters gets quantum
corrections analogous to what was found on the moduli space of the 
supersymmetric gauge theories. I focus on a simple purely bosonic
example exhibiting many interesting phenomena: massless solitons and bound
states, Argyres-Douglas-like CFTs and duality in the infrared,
and rearrangement of the spectrum of stable states from weak to strong 
coupling. At the singularities on the space of parameters, the model
can be described by a continuous theory of randomly branched polymers,
which is defined beyond perturbation theory by taking an appropriate
double scaling limit.
\end{abstract}
%
%
%
\baselineskip=18pt
\paragraph{Introduction and overview of the results}
In recent years, notable successes have been achieved in understanding
strongly coupled quantum field theories and string theories in various
space-time dimensions, by combining old heuristic ideas with the power of
supersymmetry. Perhaps most striking amongst these are the results
on four dimensional, asymptotically free, gauge theories. In \cite{SWS},
Seiberg and Witten, and Seiberg, were able to compute the exact quantum
corrections to the moduli space of vacua of various $N=1$ and $N=2$
gauge theories, using a subtle generalization of Montonen-Olive 
electric/magnetic duality \cite{MO} valid at low energy. In
\cite{Mal} a concrete proposal was made, in some particular 
supersymmetric examples, for the long-suspected string description 
of gauge theories \cite{tHooft}, particularly when the number of colors
is large. Unfortunately, these works rely heavily on very
special mathematical properties of supersymmetric theories, and it has been
impossible so far to assess their relevance to the non-supersymmetric world.

The purpose of this letter is to provide a framework where the relevance of
supersymmetric models for non-supersymmetric gauge theories can
be precisely studied. The idea is to consider a class of 
asymptotically free quantum field theories in two space-time dimensions
which are distinguished by the fact that the low energy coupling can be
changed by varying mass parameters, which thus play the r\^ ole of Higgs 
expectation values. Versions of these theories with four
supercharges show {\it quantitative} similarities with $N=2$ super
Yang-Mills \cite{Dorey}, and are thus the best possible toy models for the
four dimensional supersymmetric gauge theories. 

In this work, two main results are obtained.
First, we will see that 
{\it non-supersymmetric versions of the models discussed in \cite{Dorey}
are tractable as well, and
do continue to display qualitatively the same physics}. The quantum
corrections to the space of mass parameters $\cal M$
(which is the analogue of the 
moduli space of four dimensional gauge theories) can be computed.
Convincing evidence is found that
the physics unraveled in supersymmetric gauge theories
can be relevant in genuinely non-supersymetric models as well.
Second, we will show that one can take a double scaling limit, in the sense of the
``old'' matrix models \cite{BK}, when approaching the singularities on the
quantum space of mass parameters. This possibility was not anticipated on the
gauge theory side, and, if correct in this context,
could potentially be of great theoretical interest to
understand the field theory/string theory duality in four dimensions.

Our claims will be exemplified in this letter by studying a simple 
purely bosonic theory displaying a rich
physics akin to what was found in the four dimensional
supersymmetric gauge theories. The model is a natural
generalization of both the ${\rm O}(N)$ non-linear sigma model and the
sine-Gordon model. The fields parametrize the $N-1$ dimensional sphere
${\rm S}^{N-1}$, and the lagrangian is the sum of the standard ${\rm
O}(N)$ symmetric kinetic term and an additional interaction term  
which,  for $N=2$, reduces to the sine-Gordon potential. 
Classically, this latter term gives a mass $\sim m$ to the 
would-be Goldstone bosons.
Quantum mechanically, when $m$ is much larger that the dynamically
generated ``hadronic'' mass scale $\Lambda$, the theory is weakly coupled
and much of the physics can be read off from the lagrangian.
On the contrary, when $m\sim\Lambda$, strong quantum corrections are
expected. When $m=0$ we recover the standard ${\rm O}(N)$ non-linear
sigma model, which has a mass gap and a spectrum made of a single particle
in the vector representation of ${\rm O}(N)$ \cite{PolZad}. 

The quantum space of
parameters ${\cal M}_{\rm q}$ can be worked out by using
various techniques, including a large $N$ approximation. The
submanifold of singularities $\cal H$, defined to be the locus in $\cal M$
where some of the degrees of freedom are massless, drastically changes when
one goes from the classical to the quantum regime.
Typically, either ${\cal H}_{\rm cl}$ locally splits into two, or does
not change its shape, but in both cases the low energy physics is
different in the classical and the quantum theory. Globally we obtain a
hypersurface of singularities ${\cal H}_{\rm q}$ which delimits two
regions on ${\cal M}_{\rm q}$, one at strong coupling and the other
extending to weak coupling. On some regions of ${\cal H}_{\rm q}$, 
both a soliton, which
is in our model a generalization of the sine-Gordon soliton, and a bound
state of the elementary fields, become massless. The physics in the infrared
is then governed by a non-trivial conformal field theory, either an Ising
model or an ${\rm O}(2)$ symmetric Ashkin-Teller model. Kramers-Wannier
duality exchanges the soliton and the bound state, and inside ${\cal
H}_{\rm q}$ the notion of a topological charge is ambiguous.
The dictionary with phenomena in four dimensional supersymmetric gauge
theories is the following: our non-trivial CFTs are like Argyres-Douglas
CFTs \cite{AD}, the sine-Gordon soliton corresponds to a 't~Hooft-Polyakov
monopole and Kramers-Wannier duality is mapped onto Montonen-Olive 
duality.

Another aspect of the model is that its $1/N$ expansion can be
interpreted as a sum over topologies for randomly branched polymers.
In particular, for some values of the parameters, the model has an ${\rm
O}(N-1)$ symmetry and can be viewed as a $(N-1)$-vector model with an
infinite number of interactions (bonds involving an arbitrary number of
molecules in the branched polymers). One can approach the critical
surface ${\cal H}_{\rm q}$ by taking a suitable double scaling limit,
which on the one hand gives a description of the model in terms of
extended objects (the polymers), and on the other hand defines
non-perturbatively a continuous theory of polymers in a fully consistent
context. In particular, the model overcomes notorious difficulties with
double scaling limits in two dimensions \cite{Vec}.  
\paragraph{The model and its semi-classical properties}
We will work with a space-time of euclidean signature and write the
lagrangian of the model as
\begin{equation}
\label{lag}
L = {1\over 2} \sum_{i=1}^N  
\partial_{\mu}\phi _{i,0}\, \partial_{\mu}\phi _{i,0} + {\alpha\over 2}\,
\biggl( \sum_{i=1}^N \phi_{i,0}^2 - {1\over g_{0}^2}\biggr) + V_{\rm m}.
\end{equation}
$\alpha$ is a Lagrange multiplier implementing the constraint that the
target space is a sphere of radius $1/g_{0}^2$. Without the mass term
$V_{\rm m}$, the theory is made UV finite by simple 
multiplicative renormalizations of the fields and coupling \cite{BRE}. 
In the leading $1/N$ approximation, only the
coupling constant renormalization is needed, and we will take the
renormalized fields $\phi_i=\phi_{i,0}$, and coupling constant $g$ such that
\begin{equation}
\label{running}
{1\over g^2} = {1\over g_{0}^2} + 
{N\over 2\pi}\, \log {\mu\over\Lambda _0} = {N\over 2\pi}\, \log
{\mu\over\Lambda}\, \raise 2pt\hbox{,}
\end{equation}
where $\Lambda _0$ is the UV cut-off, $\mu$ is a sliding scale,
and $\Lambda$ is the dynamically generated mass scale 
of the theory. A mass term $V_{\rm m}$ is characterized by 
the canonical dimension of the mass parameters and the way they transform
under ${\rm O}(N)$. For example, a magnetic field has canonical
dimension 2 and transforms in the vector representation. 
Once these data are fixed, the explicit form of $V_{\rm m}$ is deduced
from renormalization theory. In our model,
the mass parameters will be taken to have canonical dimension 2 and
to transform as a symmetric traceless
rank two tensor $h_{ij}$ (they are like a tensor
magnetic field). In general, $h_{ij}$ is
multiplicatively renormalized; no renormalization is actually needed
in the leading $1/N$ expansion. By diagonalizing $h_{ij}$ we can write
\begin{equation}
\label{massterm}
V_{\rm m} = -{1\over 2}\, \sum_{i=1}^N h_i\, \phi _{i}^2.
\end{equation}
The trace part of $h_{ij}$ would correspond to a constant term in the
lagrangian, and can thus be taken to be
non zero without affecting the physics.
We will use the $N-1$ independent dimensionless physical parameters
$v_{i} = (h_{N} - h_{i})/\Lambda^{2}$ and $\cal M$ will be the $v$-space.
Using the permutation symmetry amongst the $h_i$s, we will restrict
ourselves to the region ${\cal M}^{+}$ of positive $v_i$s unless explicitly
stated otherwise.
The classical masses of the $N-1$ independent elementary fields $\phi 
_{i}$, $1\leq i\leq N-1$, are then $m_i =\Lambda\sqrt{v_{i}}$.

The model has always $N$ ${\Bbb Z}_{2(i)}$ 
symmetries $\phi _{i}\mapsto -\phi _{i}$,
and can also have additional ${\rm O}(p)$ symmetries when $p$ 
of the $v_{i}$s coincide. 

Singularities on ${\cal M}^{+}$ are found classically when some, say $p$, of
the $v_{i}$s vanish. The low energy physics is then governed 
by a standard ${\rm O}(p+1)$ non linear sigma 
model, and the $p$ massless states are the $p$ classical 
Goldstone bosons for the breaking of ${\rm O} (p+1)$ down to ${\rm 
O}(p)$. ${\cal H}_{\rm cl}$ in ${\cal M}^{+}$ thus coincide with the
hyperplanes $v_i =0$. 

Quantum mechanically, in the weakly coupled region $v_{i}\gg 1$, we can use 
semiclassical techniques to investigate the spectrum of particles 
further. It can be shown that a bound state $\phi$-$\phi$
is associated 
with the operator $\phi _{N}^{2} = 1/g_{0}^{2}-\sum _{i=1}^{N-1} 
\phi_{i}^{2}$. We will compute the mass of this state, for any 
$v_{i}$s, in the next section. One can also show that the model admits 
solitons connecting the two degenerate 
minima of the potential (\ref{massterm}) at $\phi _{N}=\pm 1/g^{2}$. 
These two minima are related by the 
spontaneously broken ${\Bbb Z}_{2(N)}$ symmetry. All the solitonic (time
independent, finite energy) solutions can be explicitly found \cite{Ferr}.
In the simple ${\rm O}(N-1)$ symmetric case where 
$v_{1}=\cdots =v_{N-1}=v\gg 1$, they correspond to trajectories
joining the two poles at $\phi _{N}=\pm 1/g^{2}$ along a meridian of the
target space sphere. They are standard 
sine-Gordon solitons of masses $M_{\rm cl}=2\Lambda \sqrt{v}/g^{2}$. 
The semi-classical quantization shows that the solitons
are particles filling multiplets of ${\rm O}(N-1)$ corresponding to
the completely symmetric traceless tensor representations. The rank 
$J$ tensor has a mass $M_{J} = M_{\rm cl} + J(J+N-3) v\Lambda ^2/
(2M_{\rm cl})$.
\paragraph{The large N approximation}
A useful technique to study our model is to use a large $N$
approximation (for a recent review, see \cite{ZJ}), where $N$ is sent to
infinity while the scale $\Lambda$ is held fixed \cite{tHooft}. The large $N$
expansion is nothing but a standard loop expansion for a non-local
effective action $S_{\rm eff}$ obtained by integrating exactly a large 
number of elementary fields $\phi _i$ from (\ref{lag}). 
For our purposes, it will be useful to keep explicitly 
the order parameter $\phi _N =\sqrt{N}\varphi$ for ${\Bbb Z}_{2(N)}$ in the
action. The effective action for large $N$ is then
\begin{eqnarray}
\label{Seff}
{S_{\rm eff}\over N} = \int\! d^2x\, \biggl\{ 
{1\over 2}\, \partial _{\mu}\varphi\, \partial _{\mu}\varphi 
+ {\alpha - h_N\over 2}\, \varphi ^2 - {\alpha\over 2Ng^2} \biggr\} +
{1\over N}\sum_{i=1}^{N-1} s\bigl[\alpha - h_i\bigr],
\end{eqnarray}
where $s[f]=(1/2)\mathop{\rm tr}\nolimits\ln (-\partial ^2 + f) -
\ln (\Lambda _0 /\mu ) \int d^2 x f/(4\pi )$ can be expanded in terms of
ordinary Feynman diagrams.

It is useful to compute, in this framework, the masses of the
particles we found previously in a semi-classical approximation.
At leading order, this is done by looking at poles in
the two-point functions derived from (\ref{Seff}).
In the ${\rm O}(N-1)$ symmetric case $v_1 =\cdots =v_{N-1}=v$, and for
$v>1$, the $N-1$ elementary fields $\phi _i$ have a mass $m_{\phi}=\Lambda
\sqrt{v}$, and the mass $m_{\rm b}$ 
of the $\phi$-$\phi$ bound state (the field $\varphi$) is a solution of
\begin{equation}
\sqrt{4\Lambda ^2 v/m_{\rm b}^2 -1}\, \ln v = 
2\arctan \biggl(1\!\Bigm/ \!\sqrt{4\Lambda ^2 v/m_{\rm b}^2 -1}\biggr).
\end{equation}
%
$m_{\rm b}$ is a monotonic function of $v$, decreasing from $m_{\rm b}
\simeq 2 m_{\phi} \bigl(1- (Ng^2)^2/32\bigr)$ for 
$v\gg 1$ to $m_{\rm b}=0$ for $v=1$. 
The mass of the ${\rm O}(N-1)$ singlet
soliton also goes to zero as $v$ goes to one because the two degenerate
minima of the effective potential derived from $S_{\rm eff}$ merge at
$v=1$. More generally, the
$\phi$-$\phi$ bound state and the lightest solitonic state will become
massless together on the critical
hypersurface $\prod_{i=1}^{N-1} v_i =1$. Near
this hyperboloid, 
the $1/N$ expansion has IR divergencies and is no
longer reliable. These divergencies are due to the fact that we are near
a critical point below the critical dimension. To describe the physics
near the critical surface, we must go beyond the $1/N$ approximation and
sum the most relevant (i.e. the most IR divergent)
contributions. This also automatically
resolves the difficulties associated with massless propagators in two
dimensions. One can show that the low energy effective lagrangian on 
the critical surface is, in the large $N$ limit,
\begin{equation}
\label{IsingLG}
L_{\rm eff}={1\over 2}\, \partial _{\mu}\phi_{N}\, \partial_{\mu}\phi_{N} +
{\pi\Lambda ^{2}\over\sum_{i=1}^{N-1}1/v_{i}}\,\, \phi_{N}^{4},
\end{equation}
where the interaction, which is proportional to $1/N$, 
must be treated exactly, since the IR divergencies compensate for the 
$1/N$ factors. We thus obtain the Landau-Ginzburg description of an 
Ising critical point. The field $\phi_{N}$ is the order operator and 
the massless soliton corresponds to the disorder operator; they are 
exchanged by Kramers-Wannier duality. Displacements on the critical surface 
are associated here with irrelevant operators in the IR.

It is important to note that the $1/N$
corrections to the equation of the critical surface itself only suffer
from mild logarithmic IR divergencies that can be handled \cite{Ferr}. 
The existence
and form of this surface can thus be reliably studied in a $1/N$
expansion. We will see that the critical surface
actually intersects with the hyperplanes $v_i=0$, and thus join with the
other sheet of the surface existing for $v_i<0$. The relevant physics
is discussed in the following section.
\paragraph{The quantum space of parameters}
There are regions in the space of parameters where it is easy to see 
that the classical and quantum hypersurfaces of singularities 
coincide. When one of the $v$s is zero, say $v_{1}=0$, and all the 
other $v$s are large, the low energy theory is an ${\rm O}(2)$ 
sigma model of small effective coupling $g_{\rm eff}$. Both 
classically and quantum mechanically, this is a massless theory, and 
thus ${\cal H}_{\rm cl}={\cal H}_{\rm q}=
\bigl\{ v_{1}=0\bigr\}$ in this region.
However, the physics are not the same: the 
classical theory of a free massless boson is replaced in the quantum case
by the non-trivial CFT of a boson compactified on a circle 
of radius $R$ with $R^{2}=4\pi/g_{\rm eff}^{2}=
\ln (\prod_{i=2}^{N-1}v_{i})$. 
If $p$ of the $v_{i}$s, $i\geq 2$, decrease, while we stay on the 
hyperplane $v_{1}=0$, the low energy theory will tend to become an 
${\rm O}(p+2)$ non linear sigma model and thus develop a mass gap. 
In the large $N$ limit, it can be shown that this transition takes place
on a surface ${\frak h}_{{\rm q},1}$ whose equation is
$\prod _{i=2}^{N-1}v_{i}=1$, $v_{1}=0$, and that the 
low energy theory on ${\frak h}_{{\rm q},1}$
is an ${\rm O}(2)$ symmetric Ashkin-Teller model 
with Landau-Ginzburg potential
\begin{equation}
\label{ATLG}
V_{\rm eff}={\pi\Lambda ^{2}\over\sum_{i=2}^{N-1}1/v_{i}}\,\, 
\Bigl( \phi_{1}^{2}+\phi_{N}^{2}\Bigr)^{2}.
\end{equation}
This model is indeed equivalent to a compactified boson for the particular
radius $R=R_{\rm KT}=2\sqrt{2}$, and the transition through ${\frak h}_{{\rm
q},1}$ is nothing but a Kosterlitz-Thouless phase transition.
A local analysis shows that ${\cal H}_{\rm q}$ joins the hyperplane $v_1=0$
orthogonally along ${\frak h}_{{\rm q},1}$, and that going from $v_1=0$ to
$v_1>0$ on ${\cal H}_{\rm q}$ is equivalent to turning on a relevant operator
which decouples one of the two Ising spins in the Ashkin-Teller model.
The other spin would decouple in the $v_1<0$ region. 
We see that the Ashkin-Teller model on ${\frak h}_{{\rm q},1}$ is made of
the coupling of the two Ising models (one for $v_1 >0$, the other for $v_1
<0$) found in the preceding section.  

\begin{figure}
\epsfxsize=15cm
\epsfbox{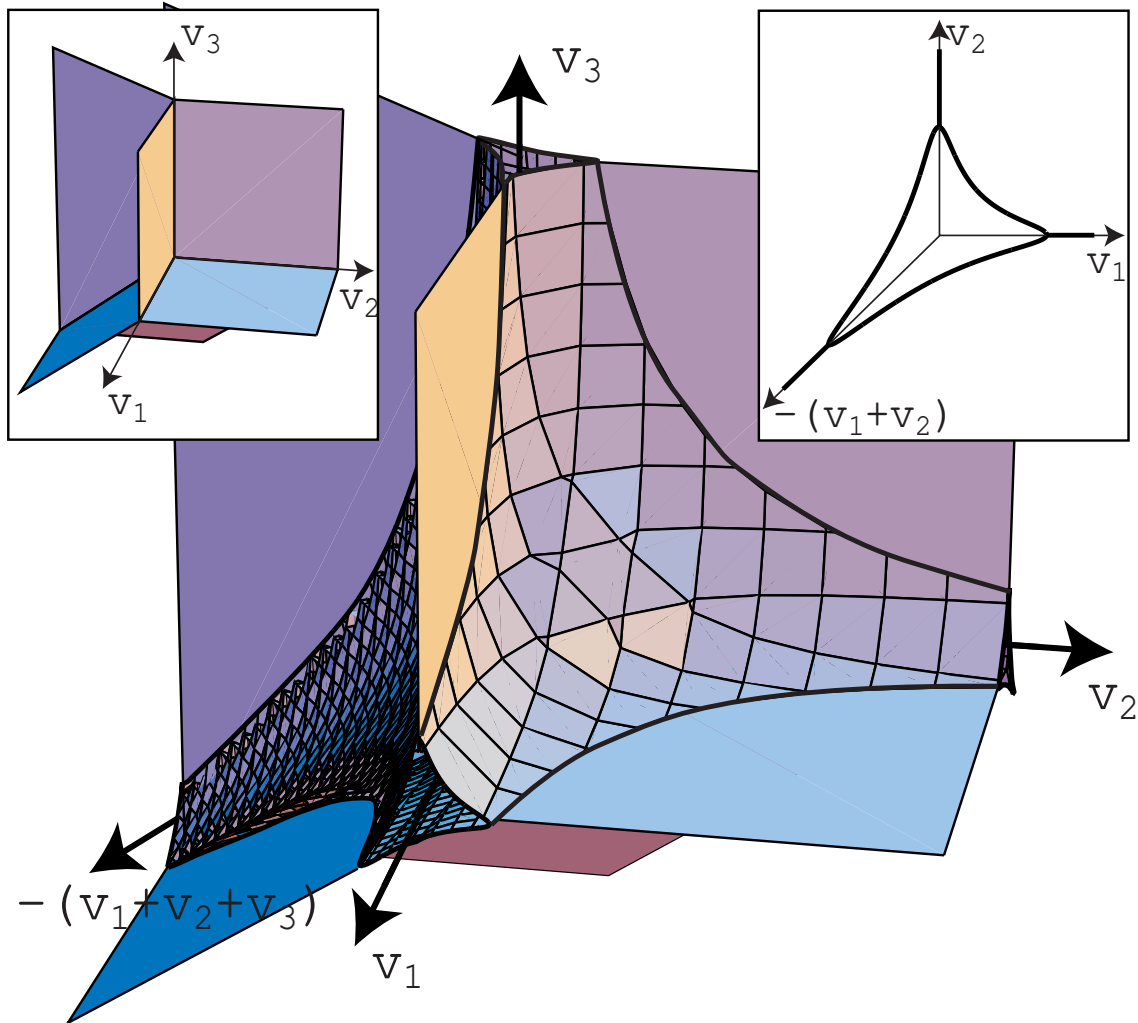}
\vfill
\caption{The classical (left inset) and quantum hypersurfaces
of singularities ${\cal H}_{\rm cl}$ and ${\cal H}_{\rm q}$
in the case $N=4$. We chose $\tilde\Lambda /\Lambda = 2/3$.
Along the four tubes extending towards infinity, one
of the fields is very massive and can be integrated out.
A section of any of the tubes thus gives the curve of singularities for
$N=3$, which is also represented (right inset). The CFTs on ${\cal
H}_{\rm q}$ can be either an Ising critical point, an Ashkin-Teller
critical point, or a compactified boson of any radius 
$R\geq R_{\rm KT} = 2\sqrt{2}$.}
\label{fig}
\end{figure}

It is possible to find a simple equation for ${\cal H}_{\rm q}$ valid for all
values of the $v_i$s, positive or negative, in the large $N$ limit:
\begin{equation}
\label{hypersurfaceH}
{\cal H}_{\rm q} :\quad
\sum_{i=1}^N \prod_{j\neq i} (r-h_j) = \Lambda ^{2(N-1)},
\end{equation}
where $r$ is the largest real root of the polynomial $\prod _{i=1}^N (x-h_i)
- \tilde\Lambda ^{2N}$. $\tilde\Lambda ^{2N}$ can in principle be determined
in terms of $\Lambda$ by a non-perturbative calculation. 
The fact that the Kosterlitz-Thouless transition occurs at a strictly
positive radius $R_{\rm KT}$ implies $\tilde\Lambda < \Lambda$.
The shape of ${\cal H}_{\rm q}$ is likely to be qualitatively reproduced by 
(\ref{hypersurfaceH}) even for small values of $N$, and we have used it
to draw Fig.\ \ref{fig}, which gives a global picture of the quantum space
of parameters of our model in the case $N=4$.
\paragraph{The double scaling limits}
On the critical surface ${\cal H}_{\rm q}$, the sum of
Feynman diagrams of a given topology (i.e. contributing at a fixed order 
in $1/N$) diverges. It is then natural to ask whether it is possible to 
approach ${\cal H}_{\rm q}$ and take the limit $N\rightarrow\infty$ in a
correlated way in order to obtain a finite answer taking into account
diagrams of all topologies. This is the idea of the double scaling limit
\cite{BK}, and for vector models the result can be interpreted as giving the
partition function of a continuous theory of randomly branched polymers
\cite{AA}.

Let us consider for example the case of the ${\rm O}(N-1)$ symmetric theory
$v_1=\cdots =v_{N-1}=v$. Near the critical point $v=1$, the effective theory
is just given by (\ref{IsingLG}) with the relevant perturbation
$(1/2)\Lambda ^2 (1-v)\, \Phi _{N}^2$. The idea (used in \cite{ZJ})
is then to eliminate the explicit $N$ dependence in the interaction term
by using a rescaled space-time variable $y=x/\sqrt{N}$,
\begin{eqnarray}
\label{dbles1}
S_{\rm eff} = \int\! d^2 y \, \biggl\{ {1\over 2}\,
\partial _{\mu}^{y} \phi _N \, \partial _{\mu}^{y} \phi _N +
{1\over 2} \Lambda ^2 N (1-v) \, \phi _{N}^2 +
\pi\Lambda ^2\, \phi _{N}^4
\biggr\}.
\end{eqnarray}
A consistent
double scaling limit can be defined to be
$N\rightarrow\infty$, $v\rightarrow 1$, with $N(v-1) - 3\ln N$ kept fixed.  
The logarithmic correction to the naive scaling comes from the fact that the
large $N$ limit is also a large UV cutoff limit in the $y$ variable,
and (\ref{dbles1}) needs to be renormalized, which is done by a simple normal
ordering. A similar non-trivial double scaling limit can be defined
near the Ashkin-Teller point. For example, for $N\rightarrow\infty$,
$v_1\rightarrow 0$, $v_2=\cdots =v_{N-1}=v\rightarrow 1$, the combinations
that must be kept fixed are $N(v-1) -4\ln N$ and $N(v-v_1-1) -4\ln N$.
Note that our theory in these limits is free of the inconsistencies found in
the standard vector models in a similar context \cite{Vec}, as can be shown
from a straightforward calculation of the 1PI effective action.
\paragraph{Other models}
It is possible to study other mass terms or/and other target spaces along
the lines of the present work. It could also be interesting to perform
lattice calculations for this class of models.
For example, a mass term $m_{ij}=m_{ji}$ of
canonical dimension one allows to obtain higher critical points. The quantum
space of parameters of a ${\Bbb C}{\rm P}^N$
model with mass terms, which has instantons, a $\theta$ angle, and
exhibit confinement at strong coupling, is also likely to display a 
rich structure. Finally, it is natural to consider supersymmetric versions
of our models. 
It turns out that the massive version of the supersymmetric
${\Bbb C}{\rm P}^N$
model shows {\it quantitative} similarities with $N=2$ super
Yang-Mills \cite{Dorey}.
A discussion of these models along with details on the present work 
will be published elsewhere \cite{Ferr,Ferr2}.

I would like to thank Princeton University for offering me matchless working
conditions.
%

%
%

\begin{references}
%
\bibitem{SWS}{N. Seiberg and E. Witten, Nucl. Phys. B {\bf 426} (1994) 19;
{\bf 430} (1994) 485(E); {\bf 431} (1994) 484,\\
N. Seiberg, Phys. Rev. D {\bf 49} (1994) 6857; 
Nucl. Phys. B {\bf 435} (1995) 129.}
%
\bibitem{MO}{C. Montonen and D. Olive, Phys. Lett. B {\bf 72} (1977) 117.}
%
\bibitem{Mal}{J. Maldacena, Adv. Theor. Math. Phys. {\bf 2} (1998) 231, \\
S.S. Gubser, I.R. Klebanov, and A.M. Polyakov, Phys. Lett. B {\bf 428}
(1998) 105,\\
E. Witten, Adv. Theor. Math. Phys. {\bf 2} (1998) 253.}
%
\bibitem{tHooft}{G. 't Hooft, Nucl. Phys. B {\bf 72} (1974) 461.}
%
\bibitem{Dorey}{A. Hanany and K. Hori, Nucl. Phys. B {\bf 513} (1998) 119,\\
N. Dorey, JHEP {\bf 11} (1998) 005,\\
N. Dorey, T.J. Hollowood, and D. Tong, JHEP {\bf 5} (1999) 006,\\
F. Ferrari, to appear.}
%
\bibitem{BK}{\'E. Br\'ezin and V.A. Kazakov, Phys. Lett. B {\bf 236} (1990)
144,\\
M.R. Douglas and S. Shenker, Nucl. Phys. B {\bf 355} (1990) 635,\\
D.J. Gross and A.A. Migdal, Phys. Rev. Lett. {\bf 64} (1990) 127.}
%
\bibitem{PolZad}{A.M. Polyakov, Phys. Lett. B {\bf 59} (1975) 79,\\
A.B. Zamolodchikov and Al. B. Zamolodchikov, Annals Phys. {\bf 120} (1979)
253.}
%
\bibitem{AD}{P.C. Argyres and M.R. Douglas, Nucl. Phys. B {\bf 448} (1995)
93.}
%
\bibitem{Vec}{P. Di Vecchia and M. Moshe, Phys. Lett. B {\bf 300} (1993)
49.}
%
\bibitem{BRE}{\'E. Br\'ezin, J.C. Le Guillou, and J. Zinn-Justin, Phys.
Rev. D {\bf 14} (1976) 2615.}
%
\bibitem{Ferr}{F. Ferrari, A model for gauge theories with Higgs fields,
PUPT-1962, LPTENS-00/28.}
%
\bibitem{ZJ}{J. Zinn-Justin, lectures at the 11th Taiwan Spring School
on Particles and Fields (1997), hep-th/9810198.}
%
\bibitem{AA}{J. Ambj\o rn, B. Durhuus, and T. J\'onsson, Phys. Lett. B {\bf
244} (1990) 403,\\
S. Nishigaki and T. Yoneya, Nucl. Phys. B {\bf 348} (1991) 787,\\
P. Di Vecchia, M. Kato, and N. Ohta, Nucl. Phys. B {\bf 357} (1991) 495.}
%
\bibitem{Ferr2}{F. Ferrari, The supersymmetric non-linear sigma model with
mass terms, to appear.}
\end{references}
\end{document}